\documentclass[journal=jpcl,manuscript=article]{achemso}
\DeclareUnicodeCharacter{0308}{} 

\usepackage{hyperref}

\usepackage{graphicx}
\usepackage{dcolumn}
\usepackage{bm}

\usepackage[utf8]{inputenc}
\usepackage[T1]{fontenc}
\usepackage{mathptmx}
\usepackage{etoolbox}
\usepackage{amsmath}
\usepackage{xcolor}
\usepackage{comment}
\usepackage{mciteplus}
\usepackage{xcolor}
\usepackage[normalem]{ulem}

\definecolor{codegreen}{rgb}{0,0.6,0}
\definecolor{darkgreen}{rgb}{0.03,0.37,0.05}

\DeclareRobustCommand*{\citen}[1]{%
  \begingroup
    \romannumeral-`\x 
    \setcitestyle{numbers}%
    \cite{#1}%
  \endgroup   
}
\DeclareMathOperator\Log{Log}

\mciteErrorOnUnknownfalse
\usepackage{xr}

\author{Edward Danquah Donkor}
\affiliation{The Abdus Salam International Center for Theoretical Physics (ICTP), Strada Costiera 11, 34151
Trieste, Italy.}
\alsoaffiliation{Scuola Internazionale Superiore di Studi Avanzati (SISSA) – via Bonomea 265, 34136 Trieste,
Italy.}

\author{Adu Offei-Danso}
\affiliation{The Abdus Salam International Center for Theoretical Physics (ICTP), Strada Costiera 11, 34151
Trieste, Italy.}
\alsoaffiliation{Scuola Internazionale Superiore di Studi Avanzati (SISSA) – via Bonomea 265, 34136 Trieste,
Italy.}

\author{Alex Rodriguez}
\affiliation{The Abdus Salam International Center for Theoretical Physics (ICTP), Strada Costiera 11, 34151
Trieste, Italy.}
\alsoaffiliation{Dipartimento di Matematica, Informatica e Geoscienze, Università degli studi di Trieste, via Valerio 12/1, 34127 Trieste, Italy}
\email{alejandro.rodriguezgarcia@units.it}

\author{Francesco Sciortino}
\affiliation{Dipartimento di Fisica, Sapienza Università di Roma, P.le Aldo Moro 5, 00185 Rome, Italy}

\author{Ali Hassanali}
\affiliation{The Abdus Salam International Center for Theoretical Physics (ICTP), Strada Costiera 11, 34151
Trieste, Italy.}
\email{ahassana@ictp.it}

\title[An \textsf{achemso} demo]{Beyond Local Structures In Critical Supercooled Water Through Unsupervised Learning\\}

\begin{document}

\begin{abstract}
The presence of a second critical point in water has been a topic of intense investigation for the last few decades. The molecular origins underlying this phenomenon are typically rationalized in terms of the competition between local high-density (HD) and low-density (LD) structures. Their identification often require designing parameters that are subject to human intervention.  Herein, we use unsupervised learning to discover structures in atomistic simulations of water close  to the Liquid-Liquid Critical point (LLCP). Encoding the information of the environment using local descriptors, we do not find evidence for two distinct thermodynamic structures. In contrast, when we deploy \emph{non-local} descriptors that probe instead heterogeneities on the nanometer length scale, this leads to the emergence of LD and HD domains rationalizing the microscopic origins of the density fluctuations close to criticality.
\end{abstract}

\begin{figure*}[!htb]
    \includegraphics[width=\textwidth]{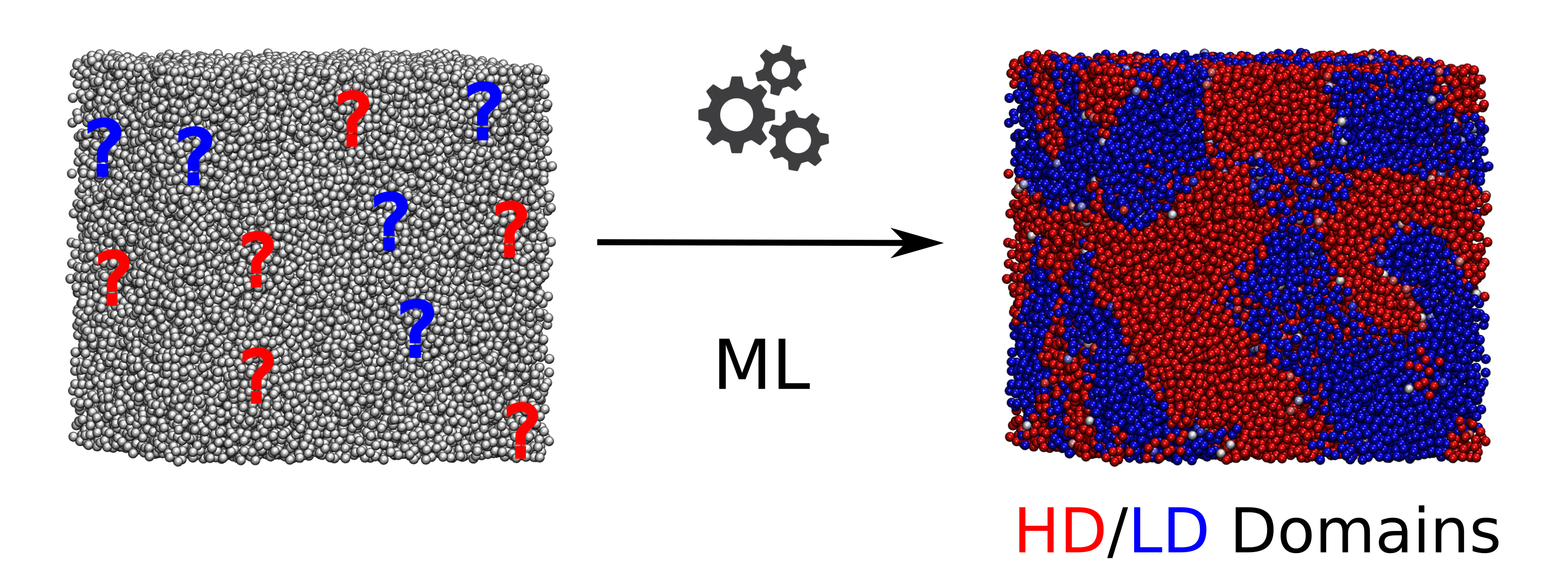}
    \captionsetup{labelformat=empty}
    \caption{TOC image}
    \label{toc}
\addtocounter{figure}{-1}
\end{figure*}

\maketitle

\newpage

\section{}

The physics of the critical behavior of matter close to phase transitions remains one of the most cherished areas of study in both experimental and theoretical physics\cite{corro2015study,shamba2012magnetocaloric,ekimov2004diamond,noguera2006theoretical}. One of the most lively areas of discussion in this regard, pertains to the microscopic origins of the complex phase diagram of water\cite{shi2018common,russo2014understanding,cuthbertson2011mixturelike,pipolo2017navigating}. Besides the rather well characterized liquid-gas critical point, a series of theoretical predictions over the last few decades have proposed the existence of another critical point - the Liquid Liquid Critical Point (LLCP) of water in the supercooled regime\cite{poole1992phase,palmer2014metastable,debenedetti2020second}. The physics underlying this criticality is thought to be one of the essential ingredients for understanding the anomalies of water.


Probing the molecular origins of this second critical point has been dominated by theoretical and numerical predictions due to the challenge of spontaneous nucleation of ice at supercooled conditions\cite{hou2018suppressing,murray2012ice,mishima2000liquid}. Just over three decades ago, Poole and co-workers demonstrated using the ST2 water model\cite{stillinger1974improved}, that deeply supercooled water showed the presence of two distinct liquid phases, with fluctuations between the two phases terminating at the LLCP \cite{poole1992phase}. Several groups have also shown in the last decade, using advanced sampling free energy calculations, that the ST2 model exhibits two distinct liquid phases\cite{liu2009low,sciortino2011study,kesselring2012nanoscale, poole2013free,   palmer2014metastable}. More recently, this has been bolstered by \emph{tour-de-force} microsecond simulations of realistic classical models of liquid water\cite{debenedetti2020second}, as well as ab initio neural network models of liquid water\cite{gartner2022liquid} that give further evidence for a LLCP scenario at least, on the numerical front. 
 On the experimental side, work on supercooled water under elevated pressures as well as pioneering sound velocity measurements appear to be breaking the boundary of the so-called \emph{no-mans land} giving strong indications of the existence of a LLCP\cite{amann2013water,kim2020experimental,amann2023liquid}.

One of the central holy-grails of understanding the possible polymorphic nature of liquid water, has been the use of locally-stable structures\cite{cuthbertson2011mixturelike,hamm2016markov,tanaka2019revealing,tanaka2020liquid,skarmoutsos2022using,foffi2024identification} which are thought to be rooted in water's unique hydrogen-bond network. Many of the anomalies in water have been rationalized in terms of a competition between two types of hydrogen-bonding structures\cite{russo2014understanding,tanaka2019revealing}. One is said to have a more ordered tetrahedral and therefore open structure, often referred to as a Low Density (LD) local configuration, and the other, disordered due to the presence of interstitial water molecules which is referred to as a High Density (HD) local configuration.

Numerous order parameters have been constructed in an attempt to identify and distinguish these two local environments\cite{errington2001relationship,chau1998new, cuthbertson2011mixturelike, tanaka2019revealing,lynden2005computational, martelli2020connection,FACCIO2022118922,montes_de_oca_structural_2020,foffi2024identification}. These order parameters have also been shown to be tightly coupled to the macroscopic density fluctuations that occur close to the critical point (CP\cite{foffi2022correlated}). Although they provide a manner in which to physically interpret the simulation data, these order parameters often require significant human intervention which necessarily involves chemical bias (and often some arbitrary cut-off in their definition). 
Furthermore, it is also not  a priori clear whether the interpretations made through these parameters are transferable across different regions of the phase diagram. 


Recently, some of us proposed a protocol that streamlines an unsupervised learning procedure for liquids and applied it to study the structure of water at room temperature\cite{offei2022high} as well as the study of the excess proton in hydrochloric acid\cite{di2023zundeig}. In brief, the method involves a three-step process.  We begin by encoding the information of local environments using local atomic descriptors computed from the Smooth Overlap of Atomic Positions (SOAP\cite{bartok2013machine}), which preserve important symmetries when comparing different molecular structures\cite{de2016comparing}. In the second step, these high-dimensional descriptors are subsequently processed through an algorithm that extracts the Intrinsic Dimension (ID)\cite{facco2017estimating} which is crucial to understand the embedding manifold of the data. In the final step, the ID is used to extract the high dimensional free energy of the system\cite{rodriguez2018computing} and identify the minima\cite{rodriguez2014clustering,d2021automatic}. For room temperature liquid water, we found a rather broad and rough landscape separated by small barriers on the order of thermal energy where the shallow minima arise from a continuum of local molecular structures that  continuously connect  the canonical low or high-density local environments\cite{offei2022high}.


In this contribution, we apply this protocol to understand fluctuations in supercooled water. Specifically, we uncover the molecular origins of critical-like fluctuations using unsupervised learning, analyzing trajectories recently reported in reference [\citen{debenedetti2020second}] connected to the presence of a second-critical point in atomistic water models. The free energy landscape constructed using local SOAP descriptors results in a single minimum despite there being macroscopic fluctuations of the global density. By systematically expanding the SOAP descriptor to include fluctuations on a length scale of up to 1 nanometer, we uncover non-local domains relevant to critical-like fluctuations in supercooled water. The free energy landscape close to the critical point evolves between the high and low-density macroscopic phases, through a complex topography which we link to collective fluctuations of chemical-based order parameters that include non-local information of the water network. 


\section{Methods}
The trajectory used for our analysis obtained from reference [\citen{debenedetti2020second}] is a $40 \mu s$ long NPT trajectory of 300 TIP4P/2005 water molecules produced using the Gromacs 5.1.4 software close to critical conditions (177K, 1751 bar). More information on the simulation conditions is detailed in the main and supplementary text of reference [\citen{debenedetti2020second}]. Additionally, we apply our analysis protocol to a large system (36424 TIP4P/2005 water molecules) in the NVT ensemble at a temperature and density of (180K, 1011.83 kg/m$^3$) in order to explore the larger length-scale fluctuations in the density.

Our unsupervised learning protocol developed in reference [\citen{offei2022high}] involves encoding local environments of molecules in a local atomic descriptor, extracting the intrinsic dimension, and constructing a high dimensional point-dependent probability density function from which the thermodynamic information can be inferred. The details of this procedure are outlined in the paragraphs below.

As indicated earlier, the first step in our analysis is to encode the water molecular environments in a local atomic descriptor. To this end, we use the Smooth Overlap of Atomic Positions (SOAP ) descriptor \cite{bartok2013machine,de2016comparing}, which preserves rotational, translational, and permutational symmetries of our molecular environments.
In brief, given an atomic environment $\chi$ around a central atom, one characterizes the local density as a sum of Gaussian functions with variance $\sigma^{2}$ centered on each of the neighbors of the central atom including the central atom itself: 

\begin{equation}
\rho_{\chi}(\mathbf{r})=\sum_{j\in \chi} \exp \left ( \frac{-\left | \mathbf{r} - \mathbf{r_{j}}\right |^2}{2\sigma ^2} \right)
\end{equation}
This atomic neighbor density can be expanded in terms of radial basis functions and spherical harmonics $Y_{lm}$ such that:
\begin{equation}
   \rho_{\chi}(\mathbf{r}) \approx \sum_{n=0}^{nmax} \sum_{l=0}^{lmax}\sum_{m=-l}^{l} c_{nlm} g_{n}(r)Y_{lm}(\theta, \phi)
 \end{equation}
where the $c_{nlm}$ are the expansion coefficients.

The number of expansion coefficients one chooses to compute is bounded by the number of radial and angular basis functions ($n_{max}$,$l_{max}$). In practice, one defines a cut-off radius ($r_{cut}$) for the atomic environment being considered. One can then define a rotationally invariant power spectrum ($\mathbf{p}$), whose elements are:
\begin{equation}
     p_{n n' l} = \pi \sqrt{\frac{8}{2l+1}}\sum_m (c_{n l m})^{\dagger}c_{n' l m}
\end{equation}
Thus, the distance between two environments $\chi$ and $\chi'$ is related to the SOAP kernel by the following expression:

\begin{equation}
 d(\chi,\chi') ={1  -  K^\mathrm{SOAP}(\mathbf{p}, \mathbf{p'})} 
\end{equation}
where:
\begin{equation}
K^\mathrm{SOAP}(\mathbf{p}, \mathbf{p'}) = \left( \frac{\mathbf{p} \cdot \mathbf{p'}}{\sqrt{\mathbf{p} \cdot \mathbf{p}~\mathbf{p'} \cdot \mathbf{p'}}}\right)
\end{equation}

Using the Dscribe package \cite{himanen2020dscribe}, the local SOAP descriptor for a water molecule $\mathbf{p}(i)$ is formed by computing the power spectrum on only oxygen species within a cutoff radius ($r_{cut}$ = 3.7 \AA) centered about each oxygen atom. 
The local SOAP descriptors encode fluctuations on the length scales around the first coordination shell. To explore non-local fluctuations, we form a \emph{glocal} SOAP descriptor by taking an average of the SOAP descriptor for each molecule and its neighbors within a distance $r_{gloc}$, given as:
\begin{equation}
    \mathbf{p}_{r_{gloc}}(i) = \frac{1}{n}\sum^{n}_{j=1} \mathbf{p}(j)
\end{equation}
Here, $n$ is the number of neighbours within a distance $r_{gloc}$ of the water molecule $i$.

We also consider $ \mathbf{p}_{global}(i)$ which is the SOAP descriptor obtained by averaging the descriptors of all molecules within a snapshot. Similar types of non-local descriptors have been previously used by Lechner and Dellago\cite{lechner2008accurate} as a means to accurately include non-local structural information in crystalline solid-state systems.

The quality, size and accuracy of the SOAP descriptors depend on the parameters that go into its definition. In particular, one needs to have a balance between the level of detail the descriptors encode and also the computational management of the datasets one uses. In this work, we compute the SOAP descriptors considering only oxygen species and with the following parameters: $n_{max}=8$ , $l_{max}=6$ and $\sigma = 1.0$ \AA \ since it offers a good balance between the level of detail of the molecular environment encoded and the size of the descriptors. In section S1 of the Supporting Information we explain how the descriptors used in the ensuing analysis and the ones that are built to include hydrogens as well as the use of smaller $\sigma$, encode similar information.

Besides the SOAP based local atomic descriptors, we are also interested in examining if and how well chemical based order parameters capture the relevant fluctuations in liquid water. Of the many order parameters, the ones of interest to us in this context were the $q_{tet}$ \cite{chau1998new,lynden2005computational}, LSI \cite{shiratani1996growth,appignanesi2009evidence,malaspina2010structural}, $d_{5}$ \cite{saika2000computer},  $\rho_{voro}$ \cite{rycroft2009voro++}, $\psi$ \cite{foffi2022correlated} and $\zeta$ \cite{russo2014understanding}. More details on these chemical order parameters are provided in section S6 of the Supporting Information as well as in the main texts of the referenced material.

In data sets with numerous dimensions, the presence of correlations among variables describing each data point suggests that the system of interest likely lies on a manifold whose dimension (the Intrinsic Dimensionality of the data set) is much lower than the embedding dimension of the data. To illustrate this, consider a set of points in three dimensions – if distributed randomly, the Intrinsic Dimensionality (ID) would be three. However, correlations between coordinates could restrict data points to lie only on the surface of a sphere, resulting in an ID of 2. 

Computing the ID is closely tied to dimensionality reduction techniques \cite{mika1998kernel,jolliffe2005principal,kruskal1978multidimensional}, where the dataset is projected into a lower-dimensional space for analysis, visualization, and interpretation. The ID denotes the minimum dimensionality in which the data can be projected by applying such techniques without significant information loss. A proper understanding of the ID guides the selection of the space to analyze system fluctuations. In our study, the ID is crucial for estimating a point-dependent density function, influencing the extraction of free energy, as elaborated later.

In this work, we employed the Two-NN estimator \cite{facco2017estimating}, a recently developed technique estimating the ID based on information from the first and second nearest neighbors of data points. This method, successfully applied to various molecular systems \cite{ansari2019spontaneously,carli2020candidate,jong2018data}, operates on the assumption that the density of a data point can be considered approximately uniform within the distance to the second nearest neighbor of a data point, demonstrating that the ratio of the second to the first nearest neighbor distances ($\mu = r_2 / r_1$) follows a specific distribution:

\begin{equation}
P(\mu) = \frac{d}{\mu ^{d +1}}
\end{equation}

Here, $d$ is the ID. Assuming independence of sampled ratios $\mu_{i}$, the ID can be estimated by maximum likelihood (other estimators are also possible) as:

\begin{equation}
d = \frac{N}{\sum_{i = 1}^{N}{\log(\mu_i)}}
\end{equation}
Where $N$ is the total number of samples in the dataset.
\newline
Using SOAP distances, we estimated the ID of the water molecule environment. The ID represents the minimum number of independent order parameters needed to describe the environment, aiding in quantifying information gained or lost with different variables \cite{camastra2016intrinsic}.

The considerations of the previous chapter have a direct impact on the reconstruction of the free energy landscape of water. To this end, understanding relevant variables characterizing structural fluctuations is essential. A common strategy is to examine probability densities along chemically-inspired variables like $q_{tet}$, LSI, and $d_{5}$ \cite{cuthbertson2011mixturelike,appignanesi2009evidence,wikfeldt2011spatially}. However, this assumes no information loss in the projection (something that cannot be strictly true if the number of variables employed is smaller than the ID of the data) and that the variable correctly encodes the process of interest. Recent techniques automatically identify important degrees of freedom \cite{tenenbaum2000global,roweis2000nonlinear} and construct free energies in high dimensions \cite{carreira2000mode,gasparotto2014recognizing,gasparotto2018recognizing,rodriguez2018computing,geissler1999kinetic}. For a detailed discussion, refer to a recent review \cite{glielmo2021unsupervised}.

In this work, we employed the Point Adaptive \emph{k}-nearest neighbor estimator (PA\emph{k}) \cite{rodriguez2018computing}, avoiding the need for projection and used successfully in studying complex molecular systems \cite{carli2020candidate,jong2018data,sormani2019explicit}. The method uses the ID as a parameter to construct a point-dependent density ($\rho_i$). This density is computed by adding a linear correction to the standard $k$-nearest neighbor estimator, where the density is $\rho_i = \frac{k_{i}}{r_{k_{i}}^{d}}$, and $k_i$'s are computed for each data point as the larger neighborhood for which the density can be considered approximately constant. The rationale is that, at constant density, the variance of the density estimation scales with $\frac{1}{\sqrt{k_i}}$ while the inclusion of regions with different densities introduces a bias term to the error, therefore the procedure controls the Bias-Variance trade-off. The point-dependent free energy is $-\Log(\rho_{i})$. Previous work shows this method accurately estimates free energy errors up to dimensions as large as 8 \cite{rodriguez2018computing}.

With point-dependent free energies, independent minima in the free energy landscape (clusters) are determined using a modified density peak clustering algorithm (DPA) \cite{d2021automatic}, an extension of the original density peak clustering \cite{rodriguez2014clustering}. 
In this procedure, cluster center candidates are chosen as those whose density is maximum within their $k_i$ neighbors. Then, the saddle points between these free energy basins are computed and the clusters are considered as coming from statistical fluctuations (and therefore merged in one) if the free energy difference between the basin minima and the saddle point is lower than $Z$ times the sum of the errors associated to these free energy estimates. The parameter $Z$ is the only free parameter in DPA clustering and can be interpreted as a measure of the statistical confidence of the clustering partition. The higher its value, the more can be one sure that the clusters are not coming from statistical fluctuations, but, at the same time, the higher the probability of losing real clusters whose statistical confidence is low due to the limited number of data points. In this work, the choice of $Z$ was made by varying it in two independently generated datasets until the clusters were consistent.

Finally, PA\emph{k} and DPA results are visualized and interpreted using the uniform manifold approximation and projection (UMAP) \cite{mcinnes2018umap}, providing a convenient way to visualize high-dimensional free energy in two dimensions \cite{becht2019dimensionality}.

To unravel the relationships between various order parameters and the macroscopic density, we use a statistical test called the Information Imbalance (IB). More details on the method is provided in reference [\citen{glielmo2022ranking}]. In brief, given a dataset with N data points and F features, one can construct different distance measures A and B using any subset of the feature space of choice, the IB is then defined as:
\begin{equation}
    \Delta \left( A \rightarrow B \right) = \frac{2}{N} \langle R^{B} | R^{A }  = 1\rangle = \frac{2}{N^2}\sum_{i,j: R_{ij}^A = 1}R_{ij}^B
\end{equation}
Where $R_{ij}^{A}$ and $R_{ij}^{B}$ are the rank matrices obtained from distances A and B respectively. Thus, $R_{ij}^{A} = 1$ if point $j$ is the first neighbour of point $i$ in space A.
With this definition, if $\Delta \left( A \rightarrow B \right) \sim 0$, then space A is predictive of B and if $\Delta \left( A \rightarrow B \right) \sim 1$, then the two spaces are unrelated.
\newline
The IB is by definition asymmetric, in the sense that if $\Delta \left( A \rightarrow B \right) \sim 0$ and $\Delta \left( B \rightarrow A \right) \sim 1$ then it means distance measure A can be used to predict B with more reliability than the reverse.

We estimate the ID of the environment around a water molecule with $\mathbf{p}(i)$, $\mathbf{p}_{r_{gloc}}(i)$ for all $r_{gloc} \in$ [3.7 \AA, 6.0 \AA, 10.0 \AA] and $\mathbf{p}_{global(i)}$. We find an ID of 5 with the purely local SOAP descriptors ($\mathbf{p}(i)$) and this decreases to 4 as we increase the radial threshold to include molecules in the whole frame indicating that the averaging enhances the correlations in the descriptor. Figure S2 in the Supporting Information shows how the ID scales as a function of the number of data points sampled from the trajectory.

\begin{figure*}[!htb]
    \includegraphics[width=\textwidth]{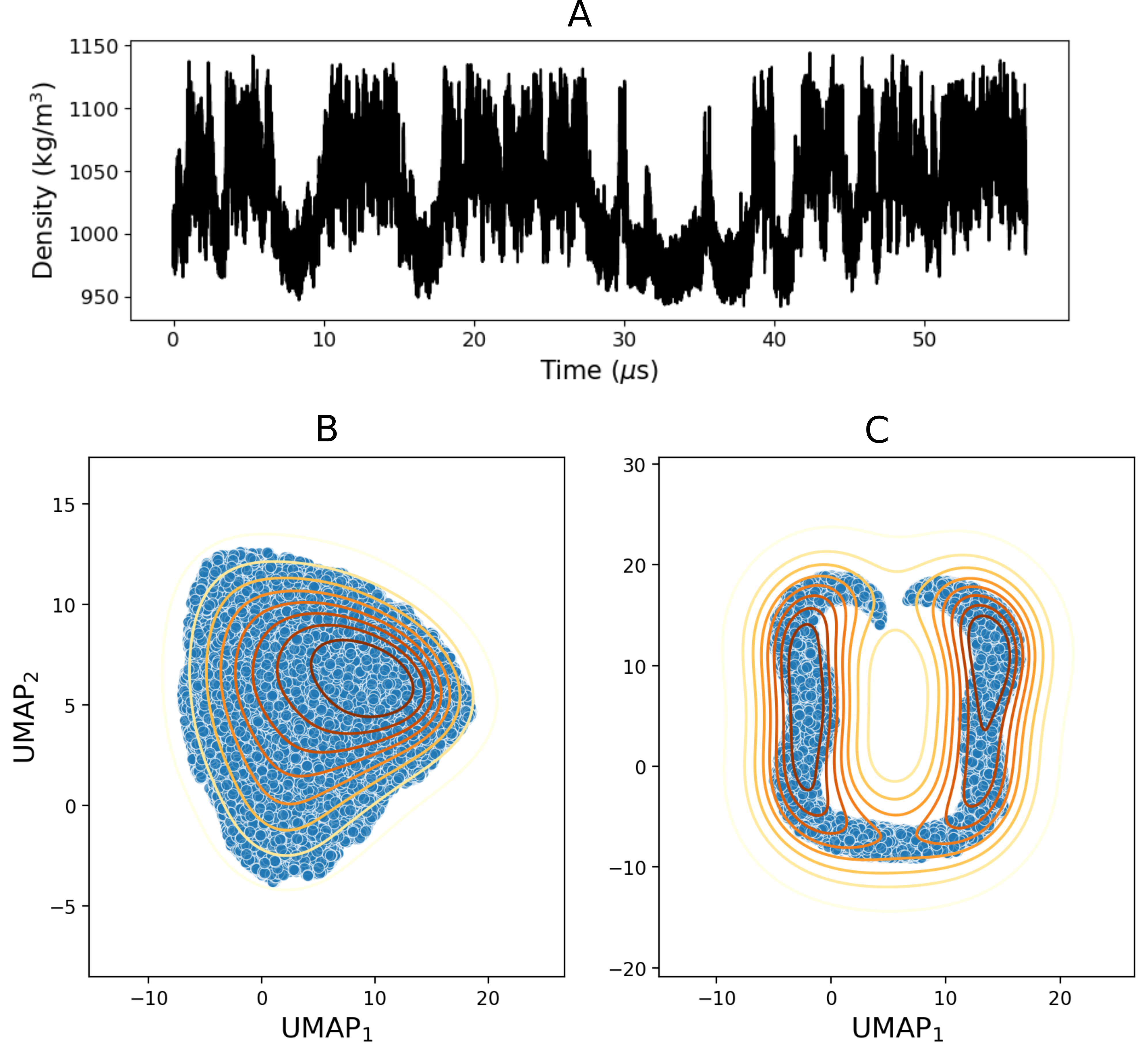}
    \caption{ (A) Critical macroscopic density fluctuations close to the critical point as obtained in reference [\citen{debenedetti2020second}]. (B) 2D UMAP representation of the data manifold obtained from the local SOAP descriptors; there is no clear separation between the two phases at the local level, as obtained from our clustering procedure and also seen from the unimodal nature of the free energy surface.  (C) 2D UMAP representation of the data manifold obtained from the global average SOAP descriptors; the two minima correspond to the Low and High macroscopic density phases obtained from our clustering procedure.}
    \label{fig:fig_1}
\end{figure*}

With the ID computed, we are now in the position to analyze the free energy landscape. 
Panel A of figure \ref{fig:fig_1} shows the time series of the simulation trajectory as reported in reference [\citen{debenedetti2020second}], where critical-like fluctuations between the HD and LD phases are observed. When using the global density as an order parameter, the underlying  free energy landscape is clearly bi-modal. Clear two-peaks distributions have been observed for several geometric and energetic order parameters, when averaged over all molecules in the system~\cite{foffi2022correlated}. 
It has also been shown that that the distributions of the same descriptors, if 
evaluated at particle level do not show a clear bi-modal character. A notable exception is
$\psi$, an indicator based on the topology of the hydrogen bond network surrounding each molecule~\cite{foffi2022correlated,foffi2024identification}.  We address the question of the onset of bimodality on crossing from local to global indicator
by performing the DPA clustering using the purely local SOAP descriptors $\mathbf{p}(i)$. Our clustering analysis reveals one cluster despite the pronounced macroscopic density fluctuations.
In panel B, we show the UMAP projection of the local SOAP descriptors in two dimensions. Confirming our clustering results, we see  that the local environments coming from both LD and HD snapshots lie in one free energy basin with no clear separation between local LD and HD environments akin to what is observed in water at ambient conditions \cite{offei2022high}. 

The microscopic origins of what we observe is likely rooted in the large heterogeneity of the local environments in both phases. Our local descriptor however, only directly encodes information of the water hydrogen bond network on the length scale of $\sim$ 3.7 \AA{}. Indeed, several previous studies have pointed to the important role of structural information beyond the second coordination shell that may be essential in understanding the differences between an HD and LD phase\cite{soper2019water,foffi2022correlated,daidone_statistical_2023}.
With this in mind, we perform the clustering with $\mathbf{p}_{r_{gloc}}(i)$. As shown in Figure S1 of the supporting information, by increasing $r_{gloc}$, the topology of the UMAP manifold starts to change and we see the emergence of two clusters (confirmed by the DPA clustering) when we average beyond the second solvation shell. Panel C of Figure \ref{fig:fig_1} shows the UMAP projection of the $\mathbf{p}_{Global}(i)$ SOAP descriptor. The clustering analysis using the global descriptors reveals two clusters which are consistent with the macroscopic HD and LD phases, further indicating that there is structural information beyond the second solvation shell that is important in distinguishing the LD and HD phases.



\begin{figure*}[!htb]
    \includegraphics[width=\textwidth]{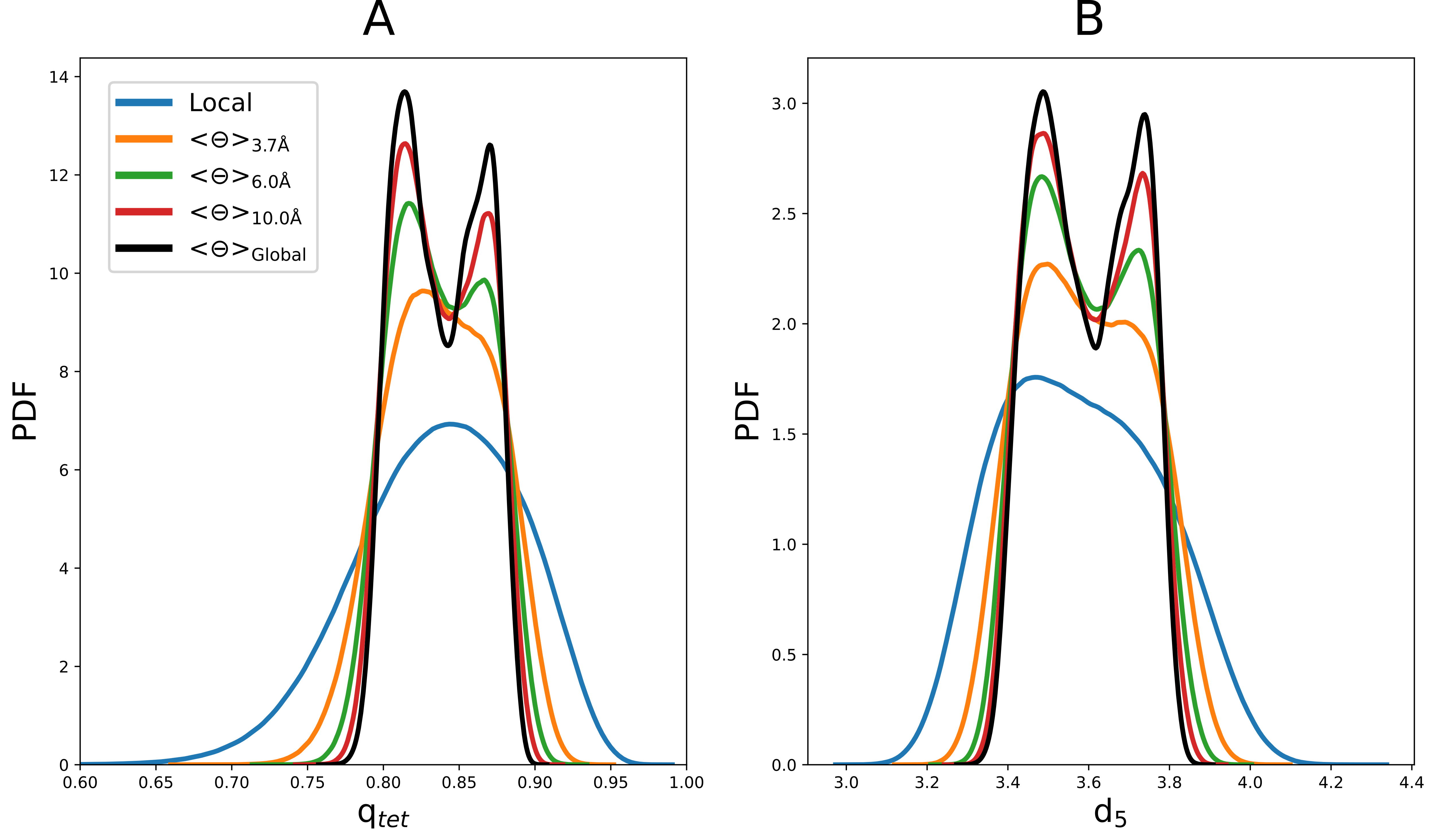}
    \caption{Evolution of the order parameters as a function of averaging. the average in the legend denotes the average within some cutoff until the global average. One can see the emergence of bimodality upon averaging beyond the second coordination shell.}
    \label{fig:fig_3}
\end{figure*}

The density plots shown in the bottom panels of Figure \ref{fig:fig_1} involves a projection of the high-dimensional SOAP features at both the local and global scale onto two UMAP coordinates which are rather difficult to interpret physically. We thus turn to examining how chemical descriptors such as tetrahedrality ($q_{tet}$) and the distance to the fifth water molecule ($d_{5}$) evolve as a function of the non-local averaging. Figure \ref{fig:fig_3} shows the distributions of the $q_{tet}$ (panel A) and $d_5$ (panel B) order parameters computed 
from the critical point trajectory in Figure \ref{fig:fig_1}. We observe that the distribution for the local order parameters is essentially unimodal with no characteristic peaks. However, upon averaging the descriptors within radial cut-offs ($<\Theta>_{r_{gloc}}$) we observe the emergence of a bi-modal structure in the distributions. Specifically, beyond $\mathrm{6}$ \AA, one peak grows at relatively low $q_{tet}$ values ($\sim$ 0.8) and hence low $d_5$ ($\sim$ 3.4 \AA) corresponding to environments sampled from the HD phase. The other peak is located at high $q_{tet}$ values ($\sim$ 0.9) and hence higher $d_5$ ($\sim$ 3.8 \AA) corresponding to average environments sampled from the LD phase. For $q_{tet}$, there is a larger proportion of environments in the LD phase consistent with what is observed with the macroscopic density\cite{debenedetti2020second} whereas this feature is much less pronounced with $d_5$.


\newpage

The emergence of bimodality in the distribution of the order parameters on longer length scales signifies that there is some correlation between them and the macroscopic density. However, the manner in which this bimodal structure develops and how strongly this reflects the density (LD vs HD) is rather sensitive to the choice of the chemical order parameter that are used. In a recent work, some of us have shown that for liquid water at room temperature, a full description of the fluctuations in the water hydrogen bond network involves the coupling of several different order parameters together\cite{donkor2023machine}. In the context of this work, we wanted to explore which order parameters and on what length-scales best probe the HD-LD density fluctuations.

\begin{figure*}[!htb]
    \includegraphics[width=\textwidth]{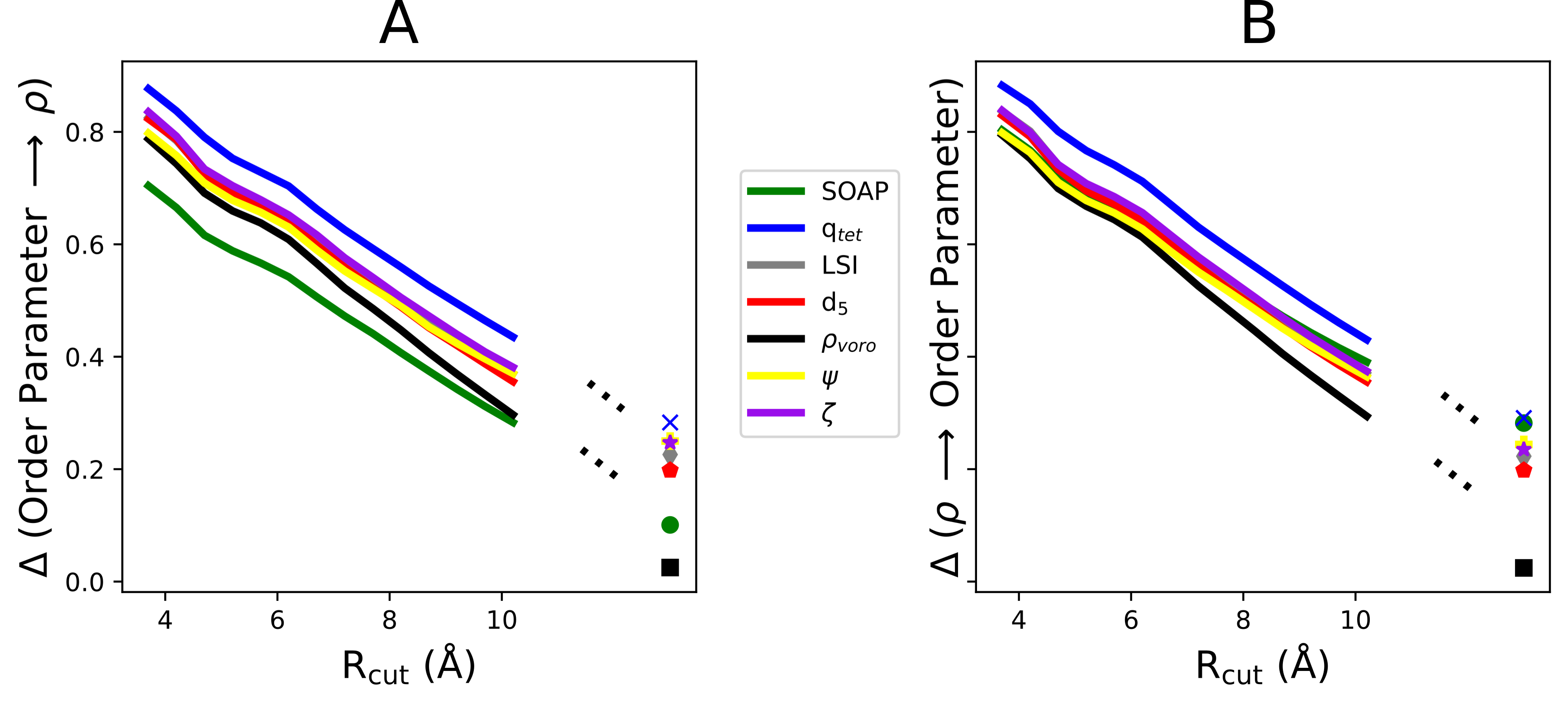}
    \caption{(A) Information Imbalance between the descriptors and the macroscopic density ($\rho$) as a function of radial averaging; we see a consistent reduction in the IB as we increase the cut-off radius for the averaging. It is noteworthy that the different descriptors reach different IB values with the whole box average, with the $\rho_{voro}$ being the most predictive of $\rho$, followed by the SOAP descriptor. On the flip side (panel B), which is the IB between $\rho$ and the descriptors, one observes a symmetry between $\rho$ and the other descriptors except SOAP, indicating that SOAP contains some information about the global average structure which $\rho$ misses.}
    \label{fig:fig_4}
\end{figure*}

To address this question, we applied the IB method to investigate the coupling between several chemical-based order parameters and the density. The IB provides a quantitative measure of how well variables such as $q_{tet}$ or $d_5$ averaged over different length scales can predict the global density and viceversa. Figure \ref{fig:fig_4} illustrates the behaviour of the IB as a function of radial averaging. We observe in panel A  that the information about the macroscopic density contained in all the descriptors  starts increasing (corresponding to low IB values) as we increase the radial cut-off for the averaging. Up to a cut-off distance of $\sim 1 nm$, which is approximately half of the whole box size, the IB reduces significantly, reaching values smaller than $\sim 0.3$ for the SOAP descriptor and $\rho_{voro}$. 

In the bottom right corners of each panel in Figure \ref{fig:fig_4}, one can see the IB value obtained for the average descriptors ($<\Theta>_{\mathrm{Global}}$)  which reaches values of $\sim 0.1$ and $\sim 0.03$ for the SOAP descriptor and $\rho_{voro}$ respectively. We note that this tight coupling between the descriptors and the macroscopic density is strictly a feature observed in sub-critical supercooled water (as we show in Figure S3 and S4 in the Supporting Information) and it is only achieved upon including structural information of up to $\sim 1 nm$ length scale as has also been discussed in reference [\citen{foffi2022correlated}]. 

The asymmetric nature of the IB allows us to also compare the information contained in the macroscopic density about the different descriptors. As seen in panel B, there is symmetric information shared between the macroscopic density and all other descriptors except for the SOAP descriptor. This is not surprising since the SOAP descriptor by nature is complete and contains information about the molecular orientations which the macroscopic density does not contain.

All in all, the preceding results builds strong evidence to a picture where the LD-HD density fluctuations cannot be described in terms of local competing structures but instead, involves clusters of at least 100 water molecules. Thus, according to our unsupervised learning protocol, the density fluctuations underlying LD-HD transitions cannot be associated with properties assigned at the single molecule level. With this picture in mind, we can revisit the UMAP projections providing more chemical interpretability. In panels A and B of Figure \ref{fig:fig_2}, we show the UMAP projection of the SOAP data in 2 dimensions, now colored with the corresponding average {$d_{5}$ ($<d_5>_{Global}$) and {$q_{tet}$ ($<q_{tet}>_{Global}$) respectively. We confirm from this, that the two density peaks (or free energy minima) emerging from our clustering correspond to the HD and LD phases since one of the peaks overlaps with $<d_5>_{Global} \sim 3.4$ \AA \  hence $<q_{tet}>_{Global} \sim 0.8$ (High Density), and the other peak overlaps with $<d_{5}>_{Global} \sim 3.8$ \AA \ and thus $<q_{tet}>_{Global} \sim 0.9$ (Low Density). 

\begin{figure*}[!htb]
    \includegraphics[width=\textwidth]{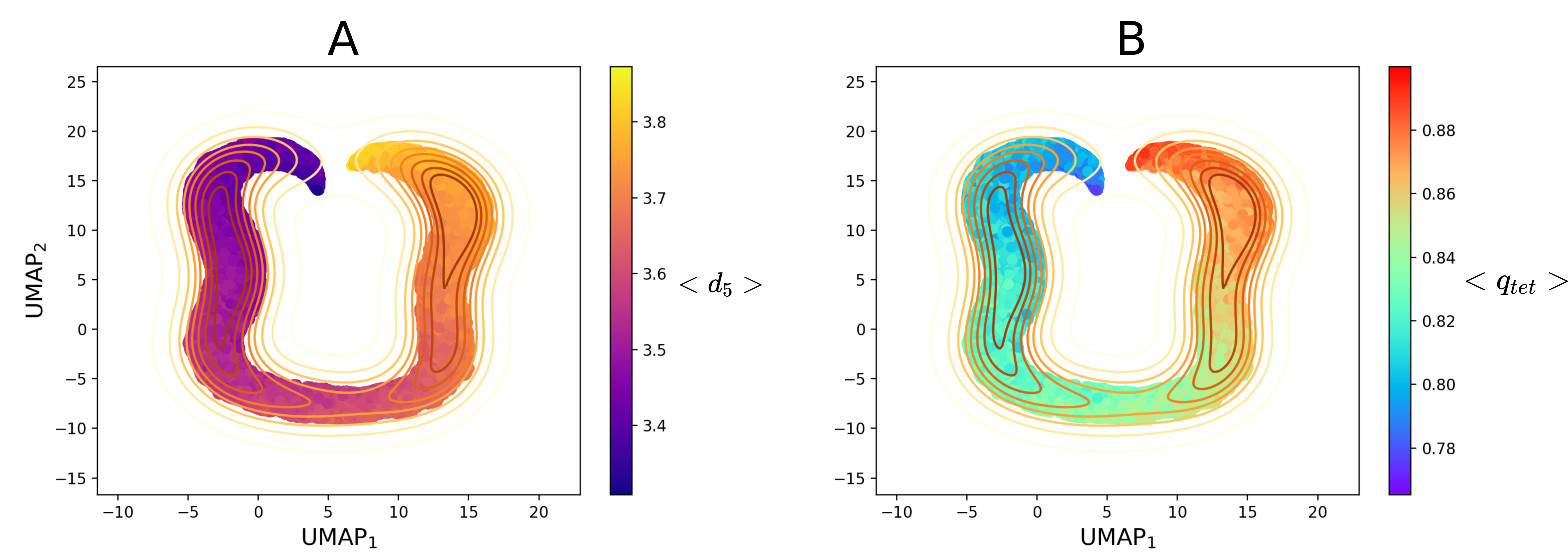}
    \caption{(A) 2D UMAP representation of the SOAP data manifold colored with the average $d_5$. The peak on the left half of this panel corresponds to low average $d_5$ while the peak on the right half corresponds to relatively high average $d_5$  and (B)  2D UMAP representation of the SOAP data manifold colored with the average $q_{tet}$. The peak on the left half of this panel corresponds to low average $q_{tet}$ while the peak on the right half corresponds to relatively high average $q_{tet}$.}
    \label{fig:fig_2}
\end{figure*}

One of the important signatures of critical behavior is the divergence in the structure factor in the low $|\vec{k}|$ limit which ultimately translates into an enhancement in long-range density fluctuations. To investigate this anomalous scattering behaviour of water close to the critical point, Debenedetti and co-workers explored the properties of the static structure factor using a large system (36424 TIP4P/2005 water molecules at sub-critical conditions run in the NVT ensemble\cite{debenedetti2020second}). Their analysis indeed shows the signature of critical behavior. A question that remains however, is how exactly one rationalizes the relationship between the non-local structures that emerge from our preceding analysis and the long-range density fluctuations.

The two clusters that have been automatically identified from the SOAP descriptors averaged on the nanometer lengthscale, (see Figure \ref{fig:fig_1} earlier) provides a protocol for classifying water molecule environments in other contexts such as those used to construct the structure factors previously described.  Using the \emph{k}-nearest neighbour classifer (see details in section S7 of the Supplemental Information), we assign water molecules in the large box simulation to either LD or HD type depending on their respective similarities.

Applying this procedure leads to the automatic identification of LD and HD domains. In the left-most panel of Figure \ref{fig:fig_5}, we show one snapshot with only oxygen atoms (for clarity) colored by the phase they have been assigned to - blue spheres represent molecules assigned as LD-like while red spheres are molecules assigned as HD-like. By visual inspection, one can see a tendency for the LD and HD water molecules to cluster together forming LD-like and HD-like domains. These domains extend over spatial distances of several nanometers and essentially percolate throughout the periodic box, and to the best of our knowledge, this is the first instance where the LD and HD domains have been identified in a completely unsupervised manner.

If the density fluctuations close to the critical point are indeed creating LD and HD domains then this implies that there should be some signature of an interfacial region forming at the boundary of the domains. One signature of this would be that water molecules close to the boundary would not be classified as pure LD or HD environments. A manner in which this can be quantified is to measure the probability of identifying either an LD or HD environment and subsequently identifying pure LD environments as those with $p_{LD} > 0.7$ and the pure HD environments as those with  $p_{HD} > 0.7$. Water molecules with $0.7 > p_{LD} > 0.4$ or $0.7 > p_{HD} > 0.4$ are then identified as those that are putatively assigned as \emph{boundary} or \emph{interfacial} points.

In the middle panel of Figure \ref{fig:fig_5}, we show the same simulation snapshot but now also coloring points that have been identified to be so-called \emph{boundary} water molecules. From visual inspection we can see how the green molecules are typically located between LD and HD domains. These findings nicely demonstrate that our procedure of agnostically identifying environments with appropriately averaged SOAP descriptors on the nanometer lengthscale leads to the emergence of LD and HD domains which are identified at the same thermodynamic state point.

In the right-most panel of Figure \ref{fig:fig_5}, we plot the PDF of the $\rho_{voro}$ order parameter constrained to the LD (blue full line), HD (red full line) and interfacial molecules (green dashed line). We note that the full distribution of $\rho_{voro}$ is unimodal and broad. However by restricting the distribution to the identified domains separately, we find that the peaks in the distributions are consistent with those associated with the LD and HD phases in the smaller box. It is also curious to observe interfacial molecules, which have $\rho_{voro}$ values peaked between the peaks of the LDL/HDL $\rho_{voro}$ distributions.

\begin{figure*}[!htb]
    \includegraphics[width=\textwidth]{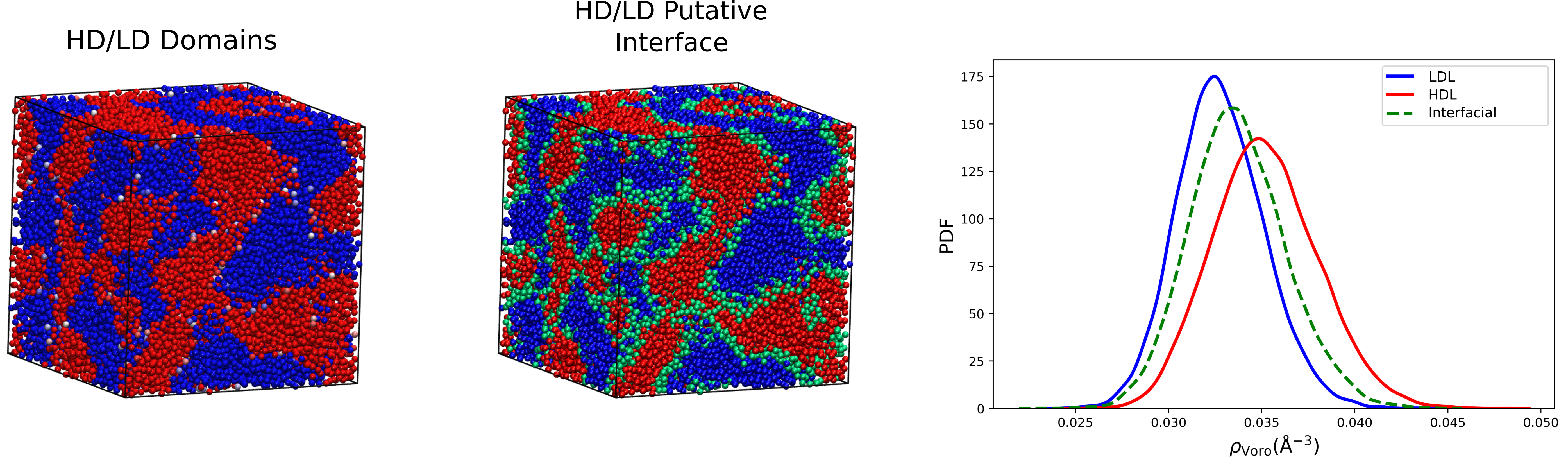}
    \caption{(Left) Snapshot of the large system colored by which density phase it was assigned to; blue for LD and red for HD. We observe the LD and HD domains extend over $1 nm$ spatial distance. (Middle) Same snapshot now coloring molecules that are found in the boundary between LD and HD domains in green. (Right) PDF of the Voronoi Density for all LD assigned molecules (Blue full line), PDF of the Voronoi Density for all HD assigned water molecules (Red full line) and the PDF of Voronoi Density values for all molecules assigned as interfacial molecules (Green dashed line). We note how the distributions are peaked towards low density, high density and intermediate density respectively, albeit with a huge overlap}
    \label{fig:fig_5}
\end{figure*}

\section*{Conclusion}

In this work, we have used unsupervised machine learning techniques to analyse data coming from molecular dynamics simulations of liquid water close to the second critical point, where one observes pronounced fluctuations of the global density between a High-Density (HD) and Low-Density (LD) liquid phase. We show that the free energy landscape in the space of local descriptors consists of one minimum despite the pronounced density fluctuations. This is rooted in the large heterogeneity of the local configurations sampled by the water molecules in both phases. However, by using descriptors that account for \emph{non-local} information of the water network, bimodality in the free energy landscape emerges. 

We further confirm the importance of \emph{non-local} information by deploying a statistical test which allows us to evaluate the strength of the mapping between different descriptors and the macroscopic density fluctuations. We find that the mapping is strongest when the descriptors are constructed to include information on approximately a nanometer length scale. Finally, armored with the \emph{non-local} HD and LD structures that emerge from our analysis, we characterize the formation of HD and LD domains that is manifested in the anomalous scattering behaviour of water close to the second critical point.

Our results bring forward important challenges in assigning and interpreting fluctuations of the hydrogen bond network in terms of single particle properties where longer range structural correlations are clearly more important. These findings should motivate more work\cite{martelli2020connection} in trying to understand the relationships between local-atomic descriptors and local-molecular chemically inspired parameters and how they change our understanding of fluctuations across the phase diagram of water. We believe our work provides a general framework for understanding water's structural and dynamic properties in other scenarios where long range correlations may be important, such as at interfaces \cite{bjorneholm2016water,das2019nature,creazzo2020enhanced} as well as under confinement \cite{di2023water,tielrooij2009effect}.

\section*{Supporting Information}
Information Imbalance between SOAP descriptors computed from different hyper-parameters, Clustering labels; from Local to Global descriptors, Intrinsic Dimension scaling; from local to global descriptors, Descriptor-Density coupling; comparison between sub-critical supercooled water and water at ambient conditions, SOAP descriptor - order parameter coupling; comparison between sub-critical supercooled water and water at ambient conditions, Description of chemical order parameters used in the main text, Test scores for the KNN model used in figure 5 of the main text.

\begin{acknowledgement}

AH thanks the European Commission for funding on the ERC Grant HyBOP 101043272. EDD thanks Alex Chen Yi Zhang for very fruitful discussions during this work.
\end{acknowledgement}

\typeout{}
\bibliography{biblio}
\newpage

\typeout{}
\pagebreak
\begin{center}
\textbf{\large Supplementary Information}
\end{center}
\setcounter{equation}{0}
\setcounter{figure}{0}
\setcounter{table}{0}
\makeatletter
\renewcommand{\theequation}{S\arabic{equation}}
\renewcommand{\thefigure}{S\arabic{figure}}
\renewcommand{\thesection}{S\arabic{section}}

\section{S1 \ \ \ Relationship between the SOAP descriptors using different environment definitions and hyper-parameters}
To make sure the SOAP power spectrum we use for our analysis contains enough information about the local environments, we use the Information Imbalance method to check the relationship between the power spectrum (using only oxygen species and $\sigma = 1.0$ \AA) and the power spectrum using oxygen and hydrogen species and also with a $\sigma = 0.25$ \AA. We find that for 4000 sampled environments $\Delta (\text{SOAP}^{\sigma = 1.0}_{\vec{\textbf{O}}} \longrightarrow \text{SOAP}^{\sigma = 0.25}_{[\vec{\textbf{O}}, \vec{\textbf{H}}]}) \sim 0.49$ while $\Delta (\text{SOAP}^{\sigma = 0.25}_{[\vec{\textbf{O}}, \vec{\textbf{H}}]} \longrightarrow \text{SOAP}^{\sigma = 1.0}_{\vec{\textbf{O}}} )\sim 0.58$.
Looking at figure 2, panel h in the original IB manuscript \cite{glielmo2022ranking}, we confirm that the two power spectra contain shared information and as such the power spectrum without hydrogens and $\sigma = 1.0$ \AA \ may be used without a significant information loss.

\section{S2 \ \ \ DPA Clustering labels: Local SOAP descriptors to \emph{glocal} SOAP descriptors}
\begin{figure*}[!htb]
    \includegraphics[width=\textwidth,]{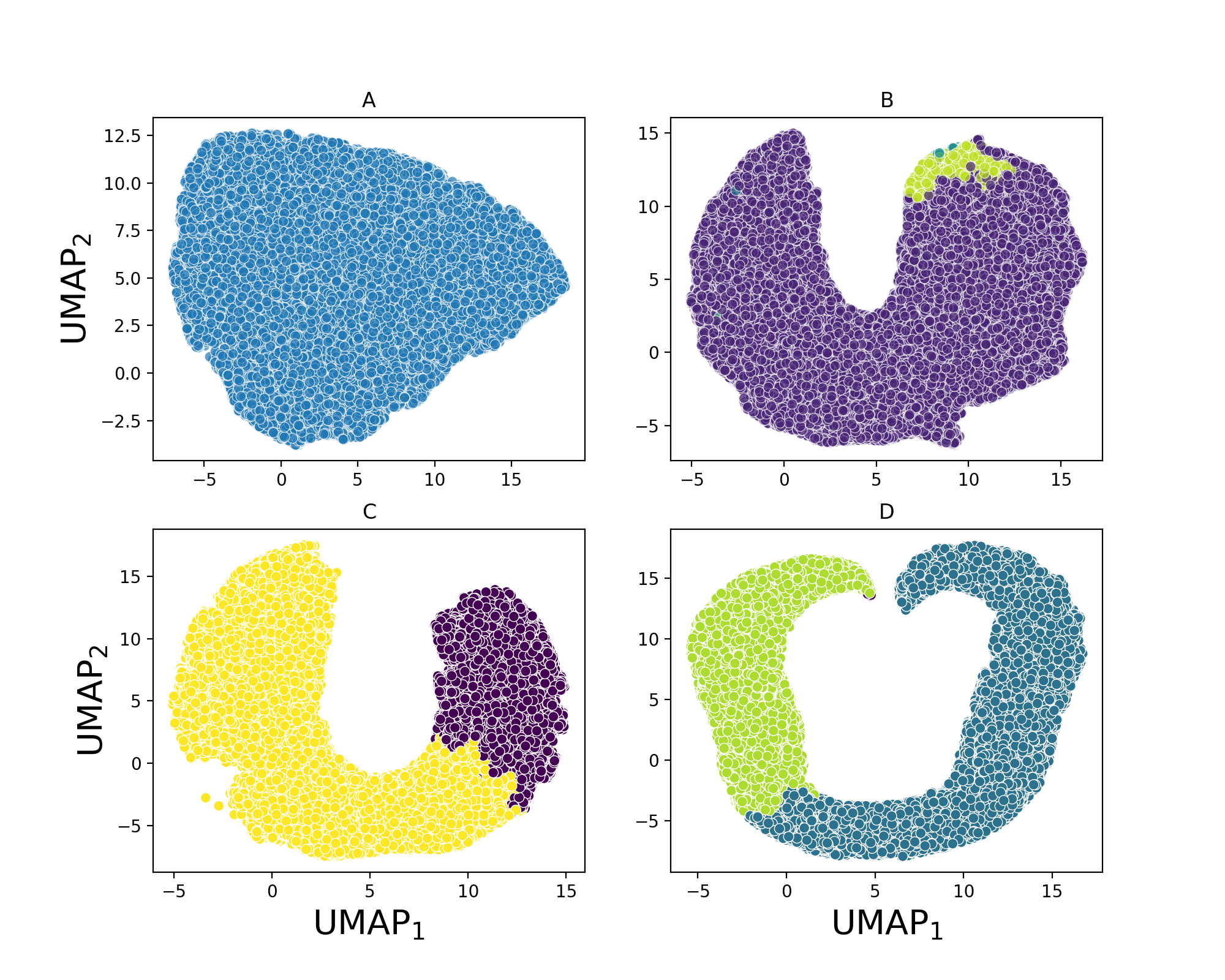}
    \caption{ Panel A,B,C and D show the 2D UMAP projection of the soap descriptor of water molecules starting from $\mathrm{SOAP}_{\vec{\textbf{O}}}$, $\mathrm{SOAP}_{\vec{\textbf{O}}_d}$, where $d\in$ [3.7 \AA, 6.0 \AA \ and 10.0 \AA] respectively. The points are colored according to the cluster assignations obtained from density peak clustering in the full SOAP space.}
    \label{umap_scaling_with_av}
\end{figure*}

\newpage
\section{S3 \ \ \ Intrinsic Dimension Scaling: from Local SOAP descriptors to \emph{glocal} SOAP descriptors}
\begin{figure*}[!htb]
    \includegraphics[width=\textwidth]{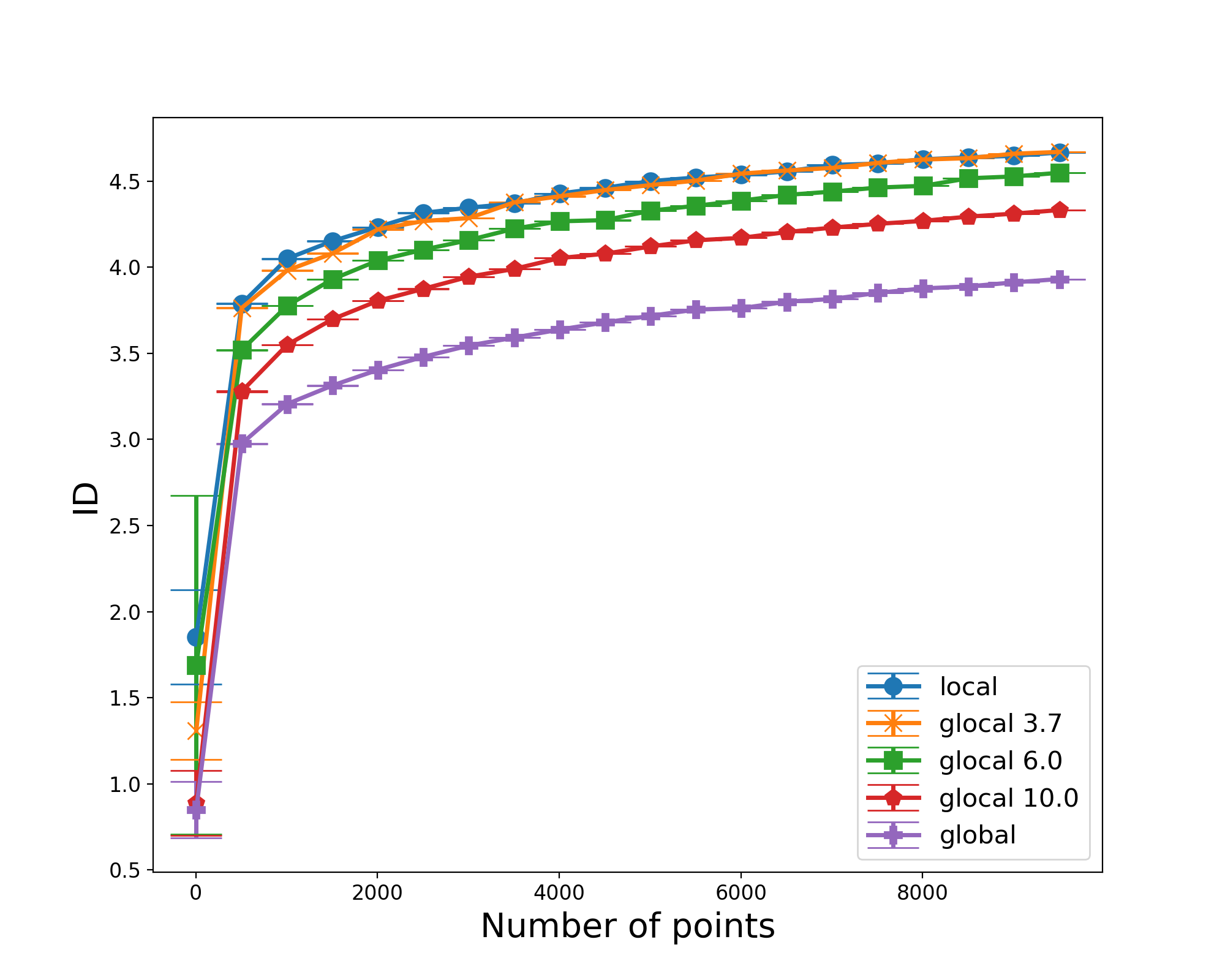}
    \caption{Scaling of the Intrinsic Dimension (ID) as a function of sampled points. The ID decreases from $\sim$ 5 to $\sim$ 4 as we increase the cut-off radius for the averaging.}
    \label{idscaling_si}
\end{figure*}
\newpage
\section{S4 \ \ \ Decriptor-Density coupling in sub-critical supercooled water in comparison to water at ambient conditions}
\begin{figure*}[!htb]
    \includegraphics[width=\textwidth]{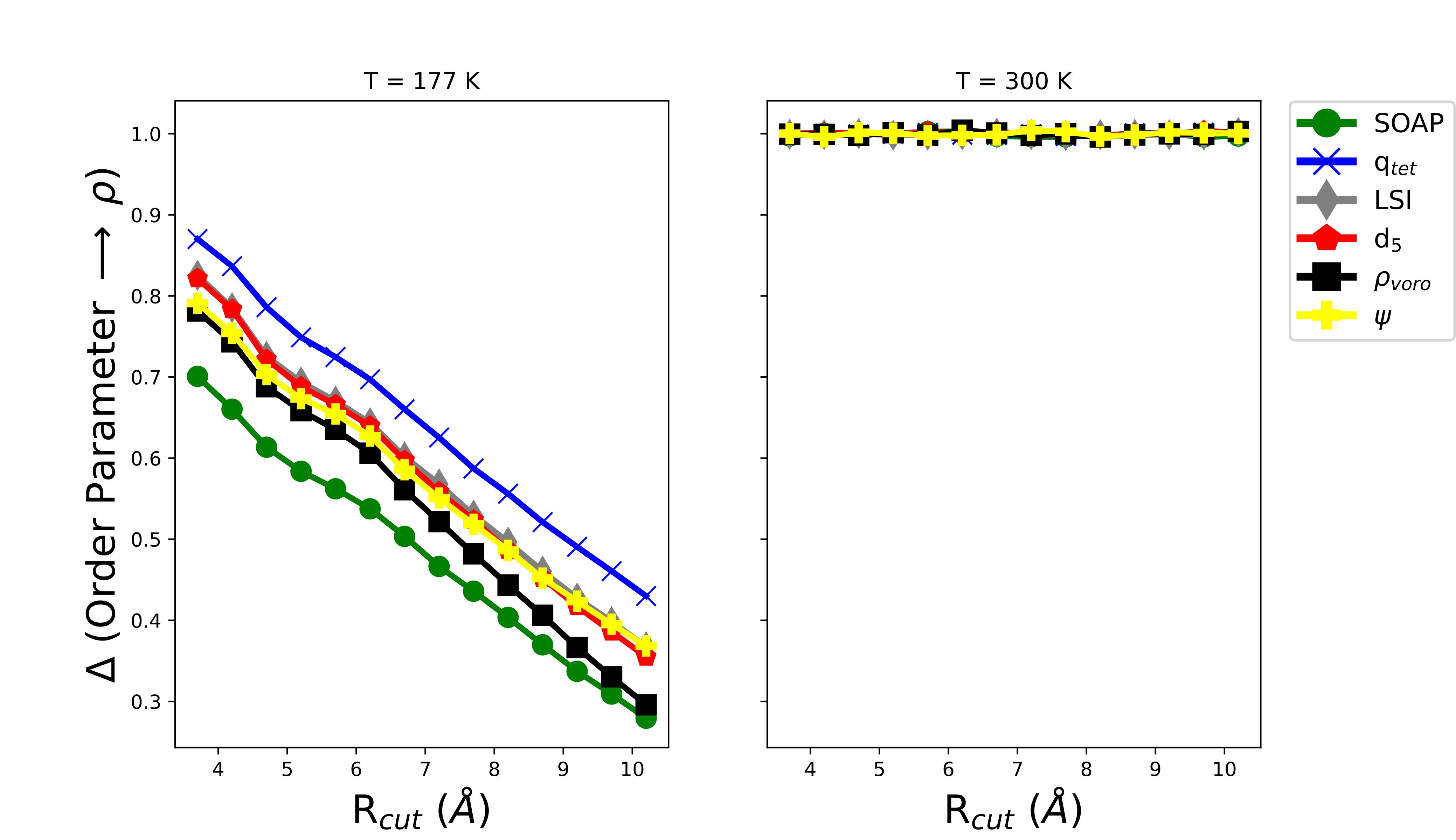}
    \caption{Information Imbalance between the different order parameters and the global density. Comparison between sub-critical supercooled water (left panel) and room temperature water (right panel). We show that the coupling between the order parameters and the global density is a feature of water in the supercooled regime as for water at ambient conditions we do not observe a decrease in the Information Imbalance with length scale of averaging.}
    \label{ib_rt_supercooled}
\end{figure*}
\newpage
\section{S5 \ \ \ Coupling Between different descriptors and the SOAP descriptor in sub-critical supercooled water in comparison to water at ambient conditions}
\begin{figure*}[!htb]
    \includegraphics[width=\textwidth]{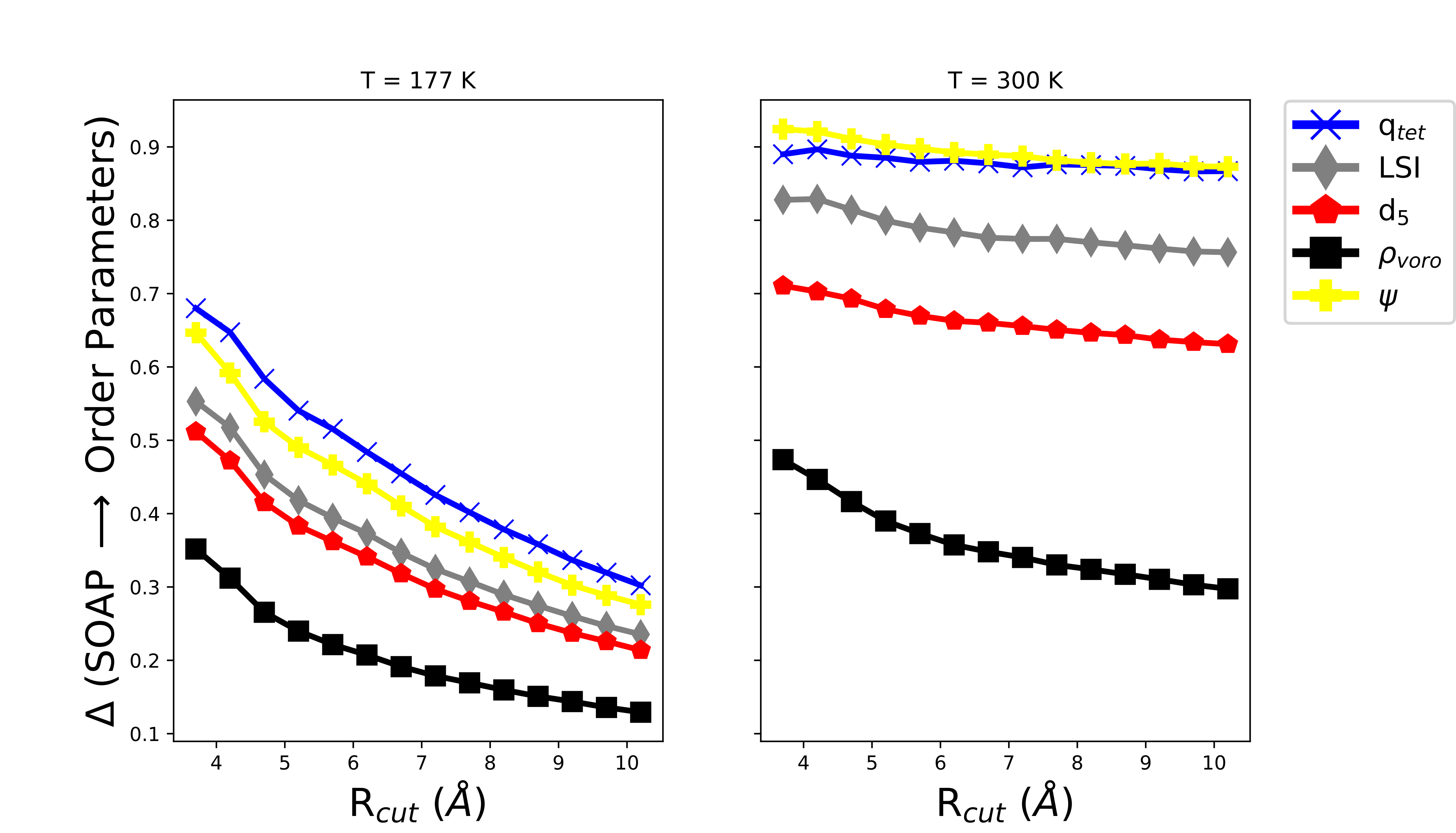}
    \caption{Information Imbalance between the SOAP descriptors and the other order parameters, comparison between sub-critical supercooled water and room temperature water. We can see that the SOAP descriptors are tightly coupled to the other descriptors at larger length scales in the supercooled regime which is not the case at ambient conditions. }
    \label{ib_rt_soap_desc}
\end{figure*}
\newpage
\section{S6 \ \ \ Chemically-Inspired Order parameters used in the study of the molecular structure of supercooled water}
\begin{itemize}
\item $q_{tet}$ measures the similarity between the first coordination  layer and a tetrahedron\cite{chau1998new,lynden2005computational}. More precisely, $q_{tet}$ is defined by the following equation: 

\begin{equation}
    q_{tet} = 1 - \frac{3}{8}\sum_{i=1}^{3} \sum_{j=i+1}^{4}( \cos(\phi_{ij}) + \frac{1}{3} )^{2}
\end{equation}

where $\phi_{ij}$ is the angle formed by the lines joining the oxygen
atom of the central water molecule to its four nearest neighbor oxygen atoms $i$ and $j$. 

\item The $d_{5}$ parameter is the distance between the fifth nearest neighbor to the central oxygen atom and reflects the extent of separation between the first and second solvation shells. A larger value of $d_{5}$ is interpreted as being a more open and locally ordered structure\cite{saika2000computer} and vice-versa, a more locally disordered and closed structure.

\item The LSI parameter was designed to distinguish environments with well-separated first and second coordination shells from those that are more disordered\cite{shiratani1996growth,appignanesi2009evidence,malaspina2010structural}. Consider the distances between the oxygen atom of a central water molecule and the $i$th
neighboring oxygen atom ordered in the following manner $r_{1} < r_{2} <...<r_{i} < r_{i+1} <...<r_{n} <3.7 \text{\AA} < r_{n+1}$. The LSI is then defined as:

\begin{equation}
    \text{LSI} =\frac{1}{n}\sum_{j=1}^{n}(\Delta(j)-\bar{\Delta})^{2}
\end{equation}
where $\Delta(j)= r_{j}-r_{j+1} $ and $ \bar{\Delta} $ corresponds to the difference and average in consecutive distances respectively.
\newline
Large values of LSI such as 0.3 correspond to structures with well separated first and second coordination shells while very low LSI values are consistent with interstitial waters between the two shells.

\item In order to measure local density variations, we computed the Voronoi density ($\rho_{voro}$), which is the inverse of the Voronoi-volume associated with a water molecule. This volume is the sum of the volume of the oxygen and two hydrogen atoms \cite{stirnemann2012communication,yeh1999orientational}. The Voronoi volume is found by performing a Voronoi tesselation on the water network.
We carry out the Voronoi tessellation using the Voro++ code \cite{rycroft2009voro++}

\item The $\psi$ descriptor, originally introduced in reference [\citen{foffi2022correlated}], is another descriptor which has been proposed to study topological arrangement of the water hydrogen bond network in supercooled conditions. More details on $\psi$ is presented in the original manuscript but in brief, $\psi$ measures the minimum physical distance between the reference oxygen atom of a water molecule and its neighbor located at a chemical distance D = 4. Where D is measured in units of number of hydrogen bonds.

\item Finally, the $\zeta$ descriptor, introduced in reference [\citen{russo2014understanding}] quantifies the separation between the first and second solvation shell around a water molecule by measuring the difference in the distance between the furthest hydrogen bonded molecule and the closest non-hydrogen bonded molecule around a central water molecule. Where two molecules are said to be hydrogen bonded using the geometric criteria by Luzar and co-workers \cite{luzar1996effect,luzar1996hydrogen}.
Several works have used this order parameter to classify water molecules into high $\zeta$ (Low Density) and low $\zeta$ (High Density) local environments \cite{russo2014understanding,shi2018microscopic,shi2018common,shi2018origin}.
\end{itemize}

\section{Classification of LD/HD environments from NVT simulations of sub-critical supercooled water}

Consider the data set consisting of N data points and D features. In our context each $i \in \{1,...,N \}$ is the 10 \AA \ \emph{glocal} SOAP descriptor of a molecule sampled from the NPT trajectory that shows strong density fluctuations.
From this data set our DPA clustering provides us with two clusters: an LD and an HD cluster. So for each data point, we have an associated label showing which phase it belongs to. 
Using these labels we perform a k-nearest neighbour classification (with k = 11) task on the 10 \AA \ \emph{glocal} SOAP descriptors computed from water environments sampled from the NVT trajectory with 36424 molecules. The aim is to identify the LD and HD domains. 
In figure \ref{fig_s6} we show that the predictive power of the k-NN model does not depend significantly on the k used. However, we use a k = 11 to get good estimates of the probability ($p$) of being assigned as an LD or HD type water.
After the classification task is carried out we obtain the probability ($p$) of being an LD or HD environment. The probability of being HD or LD are related $(p_{LD}+p_{HD}=1)$. From these probabilities we classify the core LD environments as those with $p_{LD} > 0.7$ and the core HD environments as those with  $p_{HD} > 0.7$. Then the \emph{interfacial} or \emph{boundary} molecules are labelled as those with $0.7 > p_{LD} > 0.4$ or $0.7 > p_{HD} > 0.4$. These cutoffs in the probabilities do not significantly affect the populations in the various categories of molecules.

\begin{figure*}[!htb]
    \includegraphics[width=\textwidth]{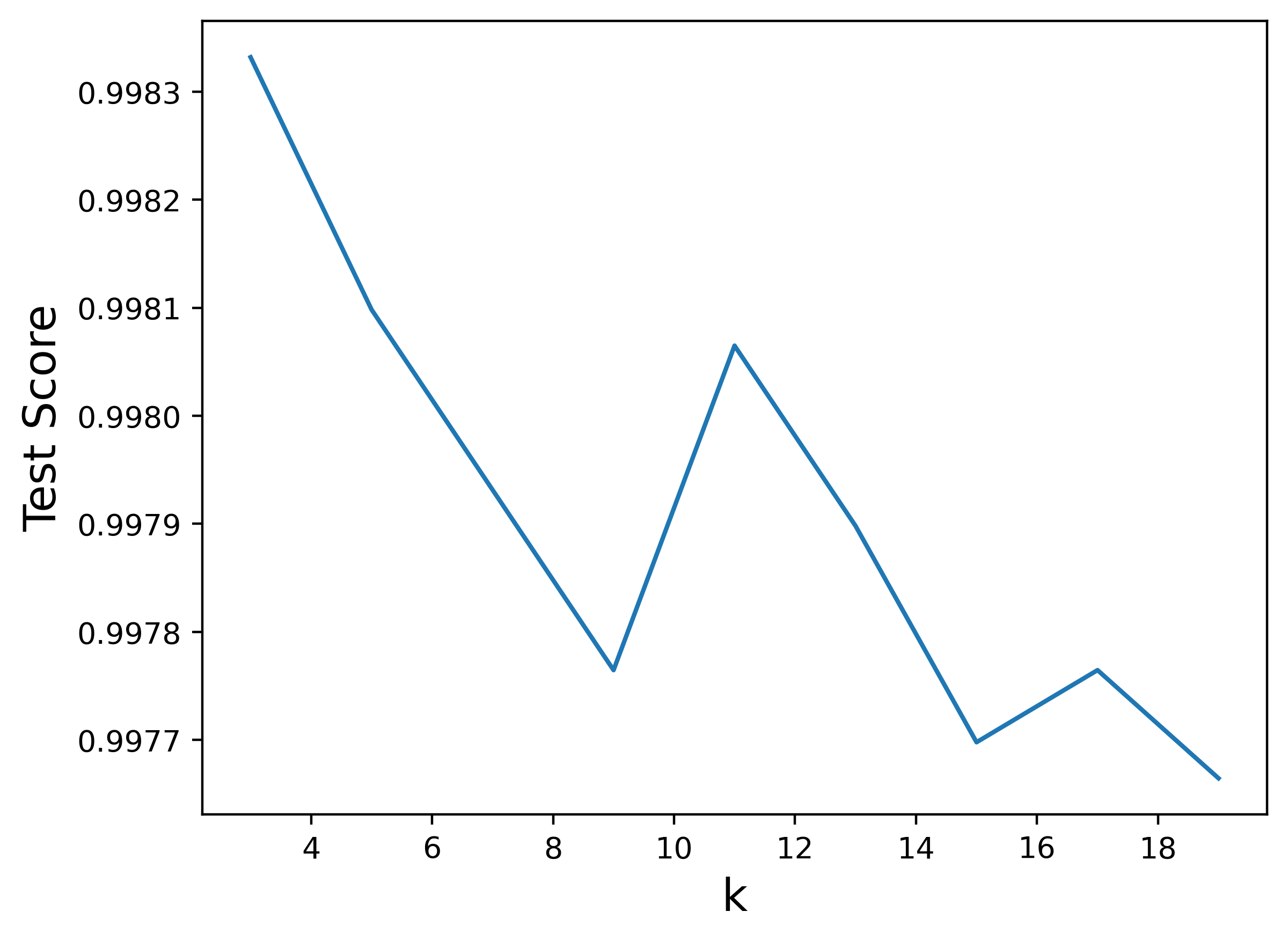}
    \caption{Test Score as a function of the number of neighbors ($k$) used. Test scores are computed by splitting our 100000 10 \AA \ \emph{glocal} SOAP descriptors from the NPT trajectory into a training and testing set. The test score is thus evaluated on the test data for several $ks$.}
    \label{fig_s6}
\end{figure*}

\end{document}